\documentclass{biophys-new}
\usepackage[utf8]{inputenc}
\usepackage{graphicx}
\usepackage[colorlinks,allcolors=cyan!70!black]{hyperref}

\usepackage[caption=false]{subfig}

\usepackage{lipsum}

\title{Statistical mechanics of an elastically pinned membrane:
Static profile and correlations}
\runningtitle{Static properties of a pinned membrane} %% For page header

\author[1,2]{Josip Augustin Jane\v{s}}
\author[1]{Henning Stumpf}
\author[1,3]{Daniel Schmidt}
\author[3]{Udo Seifert}
\author[1,2,*]{Ana-Sunčana Smith}

\runningauthor{Jane\v{s} et. al} %% For page header

 \affil[1]{PULS Group, Institut f\"ur Theoretische Physik and Cluster of Excellence: Engineering of Advanced Materials, Friedrich Alexander Universit\"at Erlangen-N\"urnberg, 91052 Erlangen, Germany}
 \affil[2]{Institut Ru\dj er Bo\v{s}kovi\'c, 10000 Zagreb, Croatia}
 \affil[3]{II. Institut f\"ur Theoretische Physik, Universit\"at Stuttgart, 70569 Stuttgart, Germany}

\corrauthor[*]{smith@physik.uni-erlangen.de}

\papertype{Article}
\begin{document}

\begin{frontmatter}

\begin{abstract}

The relation between thermal fluctuations and the mechanical response of a free membrane has been explored in great detail, both theoretically and experimentally. However, understanding this relationship for membranes, locally pinned by proteins, is significantly more challenging. Given that the coupling of the membrane to the cell cytoskeleton, the extracellular matrix and to other internal structures is crucial for the regulation of a number of cellular processes, understanding the role of the pinning is of great interest.  In this manuscript we consider a single protein (elastic spring of a finite rest length) pinning a membrane modelled in the Monge gauge. First, we determine the Green’s function for the system and complement this approach by the calculation of the mode coupling coefficients for the plane wave expansion, and the orthonormal fluctuation modes, in turn building a set of tools for numerical and analytic studies of a pinned membrane. Furthermore, we explore static correlations of the free and the pinned membrane, as well as the membrane shape, showing that all three are mutually interdependent and have an identical long-range behaviour characterised by the correlation length. Interestingly, the latter displays a non-monotonic behaviour as a function of membrane tension. Importantly, exploiting these relations allows for the experimental determination of the elastic parameters of the pinning. Last but not least, we calculate the interaction potential between two pinning sites and show that, even in the absence of the membrane deformation, the pinnings will be subject to an attractive force due to changes in membrane fluctuations.
\end{abstract} 

\end{frontmatter}

\section{Introduction}

Most living cells and a number of their internal organelles are bounded by membranes, which are composed primarily of phospholipids and proteins. The latter, in selected cases, are designed to interact with neighbouring structures thereby pinning the membrane. As such, protein complexes become spatially coordinated,  which has important consequences for the structural integrity of cells. A typical instance of such pinning is found in red blood cells, where the plasma membrane couples to the underlying spectrin network \cite{gov2004b}, although in this case additional forces associated with the soft scaffold will play a role.  \color{black}  Another example is the pinning of the membrane to stiffer scaffolds such as actin. This affects a number of cellular functions \cite{sackmann2014b}, as it  allows for the transmission of force \cite{schwarz2013}, for example, during cell adhesion. In this case, proteins such as integrins or cadherins on the plasma membrane associate into supramolecular ensembles, binding the membrane to the cytoskeleton in the cell interior and, simultaneously, to the extracellular matrix or another cell \cite{hu2017}. Similarly, inside the cell, for example on the nuclear envelope, the cytoskeleton again couples to the external nuclear membrane by nesprins, while toward the interior, protein p58 serves as a membrane attachment site for the nuclear lamina by acting as a specific receptor for lamin B \cite{Worman1988}.  All these couplings regulate the mechanical state of the cell, which in turn affects the cell motility, division rate, proliferation, mechanosensitivity, and a number of other processes \cite{hu2017}. Hence, understanding the principles of protein-mediated interactions between membranes and the surrounding scaffolds is one of the key problems in mechanobiology. 

Modeling pinned membranes,  be it the adhesion process \cite{blokhuis1999, erdmann2006, bihr2015}, in the context of the interactions with the cytoskeleton \cite{alert2015}, or the nuclear envelope \cite{lammerding2011},  requires defining the force response at the single pinning site. While different models have been used in the literature \cite{menes1997, lin2006a, netz1997}, the linear relation, where the protein attachment is described by a harmonic spring of a finite rest length, seems to capture a number of biological situations \cite{seifert2000,schmidt2012, bauer2015}.  In particular, such models have been used for more than two decades to study the interplay between the pinning sites and the forces induced by the cytoskeleton, with the assumption that the role of the membrane is merely to provide spatial coordination to the proteins. However, it is becoming more obvious that the membrane itself is not a simple spectator, but that it can act as a regulatory component \cite{perez2008, fenz2017}, since it also produces forces \cite{bell1978}. Nonetheless, because the membrane is in principle very soft, the pinning will have appreciable effects on the membrane itself.
 
Already in the early theoretical works, it was demonstrated that protein-mediated attachments of the membrane affect its shape and fluctuations \cite{dan1993, bruinsma1994,menes1997},  a fact that was used to identify binding sites in cells and vesicles \cite{smith2008,pierres2008,smith2010b}. 
 Subsequent simulations and analytical modeling showed that the mean membrane shape and roughness depend non-trivially on the instantaneous bond density \cite{lin2004b, gov2005, smith2005a, lin2006, krobath2007, reister2011,fenzbihr2011, schmidt2012}.  Alternative approaches showed, furthermore, that pinnings which experience strong frictional coupling in the membrane introduce corrections to the membrane tension  \cite{fournier2004}. Polymeric anchors, on the other hand were found responsible for the rescaling of the bending stiffness of the composite membrane in a mode-dependent fashion \cite{auth2005}. Another useful strategy relied on finding appropriate approximations to homogenize the pinning sites. As a result, a family of effective potentials that predict static properties of fluctuations were suggested in different regimes of fluctuation strength \cite{breidenich2000,merath2006, farago2008,speck2010}.    

Many studies showed that membrane fluctuations depend on the properties of the pinning itself, such as the pinning's length and mechanical stiffness \cite{bruinsma1994, seifert1997, weikl2001b, evans2003, lin2006b, hu2013, Dharan2016, fenz2017}. However, efforts to understand this coupling theoretically are scarce \cite{netz1997, speck2010, schmidt2012, bihr2012, farago2008, farago2010}. The difficulty lies in the pinning-induced coupling of plane wave modes  or spherical harmonics \cite{pecreaux2004, turlier2016}, which are otherwise independent in free membranes. The need to circumvent these technical problems led to the development of several computational approaches, which used the conveniences of  Fourier transforms and plane wave basis sets \cite{lin2006,reister2008,bihr2015}, and allowed for the numerical evaluation of mode-coupling effects \cite{fenzbihr2011}, or alternative basis sets \cite{lin2006a}. Ultimately these extensive simulations pointed to interesting many-body effects, which however could be distinguished from two-body interactions only in very limited regimes. 

In this manuscript, we provide a full analysis of static properties of a membrane pinned by an elastic spring (Fig. 1). We first calculate the static Green's function for the pinned membrane (section III), which is the working horse of analytic calculations. Given that they were not previously reported in the literature, we also provide explicit expressions for the orthonormal modes (Appendix A), and the mode coupling amplitudes for the plane wave expansion (Appendix B), both of which may be particularly useful for numerical calculations and the development of simulations, and show that they yield equivalent description as the Green's function approach. We use the Green's function to provide a comprehensive description of static properties of a pinned membrane in the full parameter range (section IV), focusing on the membrane's mean shape, fluctuation amplitude and the two-point spatial correlation function.  Besides recovering the limits known in the literature for tensionless membranes and rigid pinning, our analysis of the correlation length (section V) elucidates the interplay between the membrane rigidity and tension, the strength of the nonspecific potential and the pinning elasticity. In the final section VI, we calculate explicitly and then analyze in detail the interaction potential and the force between two pinning sites.
 
%***************************************************************************
%***************************************************************************
%***************************************************************************

\begin{figure}[t]
 \centering
 \includegraphics[width=0.5\linewidth]{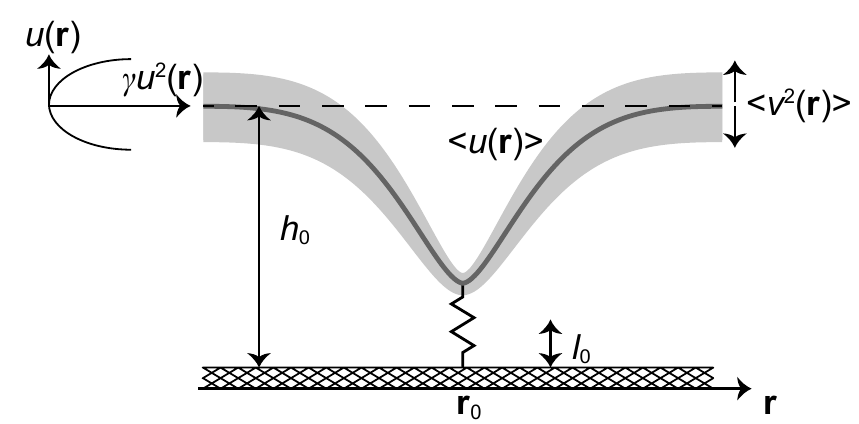}
 \caption{Mean shape $\langle u ({\bf r})\rangle$ (gray line) and the spatially dependent fluctuation amplitude $\langle v^2 ({\bf r}) \rangle$ (gray shaded area) of a membrane residing in a harmonic potential of strength $\gamma$ at $h_0$ separation from a flat substrate and pinned by an elastic spring of rest length $l_0$. 
}
 \label{fig:sketch}
\end{figure}

\section*{Methods} 
\section{Theoretical setup}

The system (Fig. 1) consists of one flexible pinning site (harmonic spring of an elastic constant $\lambda$ and rest length $l_0$, placed at the lateral position ${\bf r}_0$) that confines fluctuations of a tensed membrane (bending rigidity $\kappa$, tension $\sigma$). The membrane resides in the minimum of a harmonic non-specific potential (strength $\gamma$) at a height $h_0$ above the substrate, except near ${\bf r}_0$, where it could be displaced by the pinning. 

The membrane shape is parametrized in the linearized Monge gauge \cite{deserno2015}, such that  $u({\bf r})$ denotes deviations from the shape of a flat membrane positioned in the minimum of the nonspecific potential  along the lateral position ${\bf r}$.   Since  pinnings typically introduce membrane displacements from the minimum (order of magnitude of  1-10 nm) \cite{fenz2011, fenz2017} that are small in comparison with the correlation length of the membrane (order of 100 nm),  we use the  linearized Hamiltonian  
\begin{align}
 \mathcal H = \int\limits_A \mathrm{d}{\bf r} &\Bigg[ \frac{\kappa}{2} \left ( \nabla^2u({\bf r}) \right )^2 + \frac{\sigma}{2} \left ( \nabla u({\bf r}) \right )^2 + \frac{\gamma}{2} \left ( u({\bf r}) \right )^2  +  \frac{1}{2} \lambda\left ( u({\bf r}) - (l_0-h_0) \right )^2 \delta ({\bf r}-{\bf r}_0)\Bigg],
 \label{eq:Translated_Hamiltonian}
\end{align}
to describe the system. The first two terms in the integral on the right hand side comprise the Helfrich-Hamiltonian \cite{helfrich1978} for a bendable, pre-tensed membrane which resides  in a nonspecific potential (third term). The energetic contribution of a harmonic spring for the pinning  is represented by the fourth term which includes a delta function $\delta ({\bf r})$) positioning the pinning, as further discussed in  Supplementary Information (SI) section I. The integration goes over the projected membrane surface $A$. Here, and throughout the paper, the energy scale $k_\text{B}T$ (with Boltzmann constant $k_\text{B}$ and absolute temperature $T$), is set to unity.  The validity of this Hamiltonian has been recently discussed in detail \cite{schmidt2014}, where a reasonable agreement between numerically calculated and experimentally measured correlations and shapes has been obtained.

With $u ({\bf r}) = \langle u ({\bf r}) \rangle + v ({\bf r})$, minimization of the Hamiltonian (eq. \ref{eq:Translated_Hamiltonian}) provides the equation for the mean shape $\langle u ({\bf r})\rangle$
\begin{align}
\Bigg[\kappa \nabla^4 - \sigma\nabla^2 + \gamma + \lambda\delta ({\bf r}-{\bf r}_0)& \Bigg]  \langle u ({\bf r}) \rangle = \lambda (l_0-h_0)\delta ({\bf r}-{\bf r}_0).
 \label{eq:Mean_shape_equation}
\end{align} 

The fluctuations $v ({\bf r})$ can be obtained from diagonalizing the second variation of the  Hamiltonian, which leads to the eigenequation
\begin{align}
\left [ \kappa \nabla^4 - \sigma \nabla^2 + \gamma +\lambda \delta({\bf r}-{\bf r}_0) \right ] \psi_i ({\bf r}) = E_i \psi_i ({\bf r}),
 \label{eq:Eigenequation_eigenfunctions}
\end{align}
the latter containing the same operator as the shape equation \ref{eq:Mean_shape_equation}.
By expanding the fluctuations in these eigenmodes (see Appendix A)
\begin{align}
v({\bf r}) =  \sum_i a_i\psi_i ({\bf r}),
 \label{eq:}
\end{align}
and using the equipartition theorem
\begin{align}
 \langle a_i a_j \rangle  =  \frac{k_\text{B}T}{E_i}\delta_{ij}
 \label{eq:}
\end{align}
we find the spatial two point correlation function
\begin{align}
 \langle v({\bf r}) v({\bf r}') \rangle &=  \sum_i \frac{\psi_i ({\bf r})\psi^*_i ({\bf r}')}{E_i}.
 \label{eq:Two-point_correlation_pinned_eigenfunction}
\end{align}

We assume that the probability for membrane fluctuations with an amplitude larger than $h_0$ is small, such that these configurations will not contribute significantly to the average properties of the membrane profile. With this assumption, the details of these configurations, which would involve a non-permeable boundary at the substrate, are not important, and we can instead deal with a simpler problem in which the substrate is completely permeable to the membrane. This approximation is satisfied if the protein that pins the membrane has a finite size (larger than the fluctuation amplitude of the membrane but smaller than $h_0$), which is in experimental systems satisfied by the self-adjustment of the effective non-specific potential. Namely, if the proteins or the fluctuation amplitude  of the membrane were larger than $h_0$,  this would renormalize the non-specific potential and move the minimum away from the substrate (hence $h_0$ would be increased, and the curvature of the minimum, in our model captured by $\gamma$, would be changed), such that the required condition is recovered prior to the pinning. Practically, in the calculations this assumption is implied by having no boundary conditions on the amplitude of the membrane fluctuations.

%***************************************************************************
%***************************************************************************
%***************************************************************************

\section{Green's function approach}

\subsection{Green's function for the free membrane}

Prior to addressing the problem of a pinned membrane, it is instructive to notice that the Green's function $g_f({\bf r}\vert{\bf r}')$ for the free membrane ($\lambda$ = 0) is defined by
\begin{align}
\left [ \kappa \nabla^4 - \sigma \nabla^2 + \gamma \right ] g_f({\bf r}\vert{\bf r}') =\delta({\bf r}-{\bf r}').
\label{eq:Free_membranes_Greens_function_equation}
\end{align}
It is translationally invariant ($g_f({\bf r}\vert{\bf r}') = g_f({\bf r}-{\bf r}')$) and can be expressed as
\begin{align}
g_f({\bf r}-{\bf r}')&=\frac{1}{(2\pi)^{2}}\int_{\mathbb{R}^2} d{\bf k} \dfrac{  e^{i{\bf k}({\bf r}-{\bf r}')} }{\kappa k^4 + \sigma k^2 + \gamma}.
\label{eq:Free_membrane_Greens_function_integral}
\end{align}
The solution of the integral on the right hand side of eq. \ref{eq:Free_membrane_Greens_function_integral} is given in \cite{benhamou2011} and is a combination of modified Bessel functions of the second kind  $K_0$   
\begin{align}
g_f({\bf r}-{\bf r}') &= \frac{K_0(a_-\vert {\bf r}-{\bf r}' \vert)-K_0(a_+\vert {\bf r}-{\bf r}' \vert)}{2\pi \sigma \sqrt{1-\left(\frac{\lambda_{\text{m}}^{0}}{4\sigma}\right)^{2}}}.
\label{eq:Free_membrane_Greens_function}
\end{align}
Here,  
\begin{align}
\lambda_{\text{m}}^{0}=8 \sqrt{\kappa \gamma}
\label{eq:Lambda_m_0}
\end{align}
and the coefficients $a_{\pm}$ are given in the form
\begin{align}
a_{\pm}&=\frac{1}{\xi _0}\left[\frac{4\sigma}{\lambda_{\text{m}}^{0}}\left(1\pm\sqrt{1-\left(\frac{\lambda_{\text{m}}^{0}}{4\sigma}\right)^{2}}\right)\right]^{1/2},
\label{eq:Coefficients_a_pm}
\end{align}
with
\begin{align}
\xi_0= \sqrt[4]{\kappa/\gamma}.
  \label{eq:tensionless_correlation_length}
\end{align}
We note that the Green's function eq. \ref{eq:Free_membrane_Greens_function} is real even if $a_{\pm}$ are complex numbers. 

 As for any quadratic integral kernel, the Green's function $g_f({\bf r}-{\bf r}') $, and  respectively $g_f(0)$ are associated with the spatial correlation function $\langle v_f({\bf r}) v_f({\bf r}') \rangle$ and the mean square fluctuation amplitude $\langle v^{2}_f({\bf r}) \rangle$ of the free membrane, initially calculated by several groups  \cite{helfrich1984, lipowsky1991a, lipowsky1995, seifert1997}.  The later is commonly denoted by $1/\lambda_{\text{m}}$ \cite{bihr2012,bihr2015, fenz2017}. Hence,

\begin{align}
 \label{eq:fluctuation_amplitude_Green}
g_f(0)=\frac{1}{{\lambda}_{m}} = \frac{\text{arctan}\left(\sqrt{\left(\frac{\lambda_{\text{m}}^{0}}{4\sigma}\right)^{2}-1}\right)}{2\pi\sigma\sqrt{\left(\frac{\lambda_{\text{m}}^{0}}{4\sigma}\right)^{2}-1}},
\end{align}
which for a tensionless case  \cite{raedler1995} simplifies to
\begin{align}
 g_f(0)_{|\sigma=0}& = \frac{1}{\lambda_{\text{m}}^{0}}.
  \label{eq:tensionless_green_free}
\end{align}
Under this conditions, eq. \ref{eq:Free_membrane_Greens_function} adopts the well-known form  \cite{bruinsma1994,lipowsky1995}  
\begin{align}
 g_f({\bf r}-{\bf r}')_{|\sigma=0}&= -\frac{4}{\pi \lambda_{\text{m}}^{0}} \text{kei}_0 \left ( \frac{\vert {\bf r}-{\bf r}'\vert}{\xi_0} \right ),
  \label{eq:tensionless_green_free}
\end{align}
with $\text{kei}_0$ being the Kelvin function and $\xi _0$ being the lateral correlation length of the free tensionless membrane given by eq. \ref{eq:tensionless_correlation_length}.

%***************************************************************************
%***************************************************************************
%***************************************************************************
\subsection{Green's function for the pinned membrane}

The Green's function  $g({\bf r}\vert{\bf r}')$ providing the response of a membrane at the position ${\bf r}$ due to a disturbance at the position ${\bf r}'$  is defined as
\begin{align}
\left [ \kappa \nabla^4 - \sigma \nabla^2 + \gamma +\lambda \delta({\bf r}-{\bf r}_0) \right ] g({\bf r}\vert{\bf r}') = \delta ({\bf r}-{\bf r}').
 \label{eq:Greens_function_equation}
\end{align}
With the use of eq. \ref{eq:Free_membranes_Greens_function_equation}, eq. \ref{eq:Greens_function_equation} can be recast as
\begin{align}
\left [ \kappa \nabla^4 - \sigma \nabla^2 + \gamma \right ] \left[ g({\bf r}\vert{\bf r}')+ \lambda g_f({\bf r}\vert{\bf r}_0)g({\bf r}_0\vert{\bf r}') - g_f({\bf r}\vert{\bf r}') \right ] = 0,
\label{eq:Greens_function_algebraic_equation}
\end{align}
which can be generally valid only if the second bracket identically vanishes. Consequently,
\begin{align}
 g({\bf r}\vert{\bf r}') = g_f({\bf r}\vert{\bf r}') - \lambda g_f({\bf r}\vert{\bf r}_0)g({\bf r}_0\vert{\bf r}').
\label{eq:Greens_function_algebraic_equation}
\end{align}
Setting ${\bf r}$=${\bf r}_0$ in eq. \ref{eq:Greens_function_algebraic_equation} provides
\begin{align}
 g({\bf r}_0\vert{\bf r}')  = \frac{g_f({\bf r}_0\vert{\bf r}')}{1+\lambda g_f({\bf r}_0\vert{\bf r}_0)} = \frac{\lambda_{\text{m}}}{\lambda +  \lambda_{\text{m}}} g_f({\bf r}_0\vert{\bf r}'), 
\label{eq:system_of_equations_for_r_j_Greens_function}
\end{align}
which, upon reinsertion into eq. \ref{eq:Greens_function_algebraic_equation}, gives rise to the Green's function for the pinned membrane 
\begin{align}
 g({\bf r}\vert{\bf r}') &= g_f({\bf r}-{\bf r}') - \frac{\lambda\lambda_{\text{m}}}{\lambda+\lambda_{\text{m}}}g_f({\bf r}-{\bf r}_0)g_f({\bf r}_0-{\bf r}').
\label{eq:Greens_function_of_a_pinned_membrane0}
\end{align}
Although $g({\bf r}\vert{\bf r}')$ is comprised of the translationally invariant $g_f({\bf r}-{\bf r}')$, it itself is not generally translationally invariant.

\subsection{Representing shape and fluctuations}

By construction, $g({\bf r}\vert{\bf r}_0)$ differs only by a prefactor from the solution of the shape equation \ref{eq:Mean_shape_equation}
\begin{align}
 \langle u ({\bf r}) \rangle &=\lambda (l_0-h_0) g({\bf r}\vert{\bf r}_0).
\label{eq:Mena_shape_by_Greens_function_convolution}
\end{align}
Combining  eqs. \ref{eq:system_of_equations_for_r_j_Greens_function} and \ref{eq:Mena_shape_by_Greens_function_convolution} gives the mean shape
\begin{align}
 \label{eq:Mean_profile}
 \langle u ({\bf r}) \rangle = \frac{\lambda \lambda_{\text{m}}}{\lambda + \lambda_{\text{m}}} (l_0-h_0) g_f({\bf r}-{\bf r}_0).
\end{align}
As shown previously \cite{schmidt2012,bihr2012}, in the tensionless case combining eqs. \ref{eq:tensionless_green_free} and \ref{eq:Mean_profile} yields
\begin{align}
 \label{eq:Mean_profile__tensionless}
 \langle u ({\bf r}) \rangle _{|\sigma=0} = -\frac{4}{\pi}\frac{\lambda}{\lambda + \lambda^{0}_m} (l_0-h_0) \text{kei}_0 \left ( \frac{\vert {\bf r}-{\bf r}_0\vert}{\xi_0} \right ),
\end{align}
which  is a function of the kei function,  as expected for the differential operator of the shape equation that is bilaplacian plus a constant \cite{costa1985, chen2003}. 
In the limit of an infinitely stiff pinning $\lambda\to\infty$, eq. \ref{eq:Mean_profile__tensionless} reproduces the result obtained in \cite{bruinsma1994} .

By comparing the bilinear expansion of the Green's function in the eigenfunctions $\psi_j$ (eq. \ref{eq:Eigenequation_eigenfunctions})
\begin{align}
g({\bf r}\vert{\bf r}') = \sum_j \frac{\psi_j ({\bf r})\psi^*_j ({\bf r}')}{E_j},
 \label{eq:Operators_Greens_function_expansion in_the_operators_eigenfunctions}
\end{align}
and eq. \ref{eq:Two-point_correlation_pinned_eigenfunction}, we find
\begin{align}
\langle v ({\bf r})  v ({\bf r}') \rangle = g({\bf r}\vert{\bf r}'),
 \label{eq:Greens_function_as_a_two_point_correlation_function}
\end{align}
where the factor $k_\text{B}T=1$ on the right hand side is implicit. Hence,
\begin{align}
 \langle &v({\bf r}) v({\bf r}') \rangle = g_f({\bf r}-{\bf r}')-\frac{\lambda\lambda_{\text{m}}}{\lambda+\lambda_{\text{m}}}g_f({\bf r}-{\bf r}_0)g_f({\bf r}_0-{\bf r}').
\label{eq:Spatial_correlations_Greens}
\end{align}
Naturally, by setting ${\bf r}'={\bf r}$ in eq. \ref{eq:Spatial_correlations_Greens} we obtain the fluctuation amplitude  
\begin{align}
 \langle v^{2}({\bf r}) \rangle= \frac{1}{\lambda_{\text{m}}}-\frac{\lambda \lambda_{\text{m}}}{\lambda + \lambda_{\text{m}}} g^{2}_f({\bf r}-{\bf r}_0),
\label{eq:fluctuation_amplitude_pinned}
\end{align}
with
\begin{align}
 \langle v^{2}({\bf r}_0) \rangle= \frac{1}{\lambda + \lambda_{\text{m}}}.
\label{eq:fluctuation_amplitude_pinned_at_the_pinning}
\end{align}
The same result can be obtained by calculating the eigenfunctions $\psi_j ({\bf r})$ for a system with a single pinning (see Appendix A.1) 
\begin{align}
&\psi_{m}({\bf r},q) =\frac{i^me^{im\phi}}{\sqrt{  \left(1+\delta_{m0}\left(\Pi(q)\right)^{2}\right)}}\left[ J_m(q r) + \delta_{m0} \Pi(q) \left ( Y_m(q r) + \frac{2}{\pi} K_m\left(\sqrt{q^{2}+\frac{\sigma}{\kappa}} r\right) \right) \right],
\label{eq:Eigenfunctions} 
\end{align}
and using eq. \ref{eq:Operators_Greens_function_expansion in_the_operators_eigenfunctions} to obtain the Green's function (see Appendix A.2) .

%***************************************************************************
%***************************************************************************
%***************************************************************************

\section*{Results} 
\section{Properties of the mean shape and the correlation function} 

While the previous sections reveal the formal framework describing the effect of the pinning on the fluctuations of the membrane, several results warrant further discussion. Specifically, inserting the solution for the mean shape eq. \ref{eq:Mean_profile} into the Hamiltonian eq. \ref{eq:Translated_Hamiltonian} determines the total elastic energy of the average configuration of the system (pinning and membrane) 
\begin{align}
 \mathcal H\left[\langle u ({\bf r}) \rangle \right] = \frac{1}{2} \frac{\lambda \lambda_{\text{m}}}{\lambda + \lambda_{\text{m}}}(h_0-l_0)^2 \equiv  \frac{1}{2} \mathcal K (h_0-l_0)^2
 \label{eq:deformationenergy}
\end{align}
Equation \ref{eq:deformationenergy} shows that the deformation energy increases quadratically with the height separation between the free membrane and the pinning, while it vanishes for $h_0=l_0$ as described previously \cite{bruinsma2000, schmidt2012}. The prefactor $\mathcal K$ is an effective spring constant made up of two "springs" (the membrane and the pinning) connected in series, with $\lambda_{\text{m}}$ being the membrane spring constant. From this point of view, $ \mathcal K$ can be seen as the effective elastic constant of the system \cite{bihr2012,bihr2015, fenz2017}. 
  
The quadratic nature of eq. \ref{eq:deformationenergy} is consistent with the quadratic form of the Hamiltonian eq. \ref{eq:Translated_Hamiltonian} and the "local" nature of the pinning. A further consequence is the linear relation between the mean shape and the correlation function from the pinning site 
\begin{align}
 \langle u ({\bf r}) \rangle &=-\lambda (h_0-l_0) \langle v({\bf r}) v({\bf r}_0) \rangle,
\label{eq:Mena_shape_rescaled_correlation}
\end{align}
which emerges by inspection of eqs. \ref{eq:Mena_shape_by_Greens_function_convolution} and \ref{eq:Greens_function_as_a_two_point_correlation_function}. Here, the spatially independent prefactor has a form of a force on a harmonic spring. As a result, both the shape and the correlation function have the same features but due to a minus sign on the left hand side of eq. \ref{eq:Mena_shape_rescaled_correlation}, the trends are opposite. For instance, the well-documented overshoot of the membrane shape \cite{bruinsma1994, schmidt2012, chen2003} at distances of a couple of correlation lengths from the pinning is reflected in the anticorrelations in the same range (Fig. 2). Likewise, the displacement of the mean shape from the minimum of the non-specific potential increases with the increased pinning stiffness $\lambda$ (Fig. 2a), while the amplitude of the pinning site correlation $\langle v({\bf r}) v({\bf r}_0) \rangle$ decreases (Fig. 2b).

\begin{figure}[t]
 \centering
  \includegraphics[width=1\linewidth]{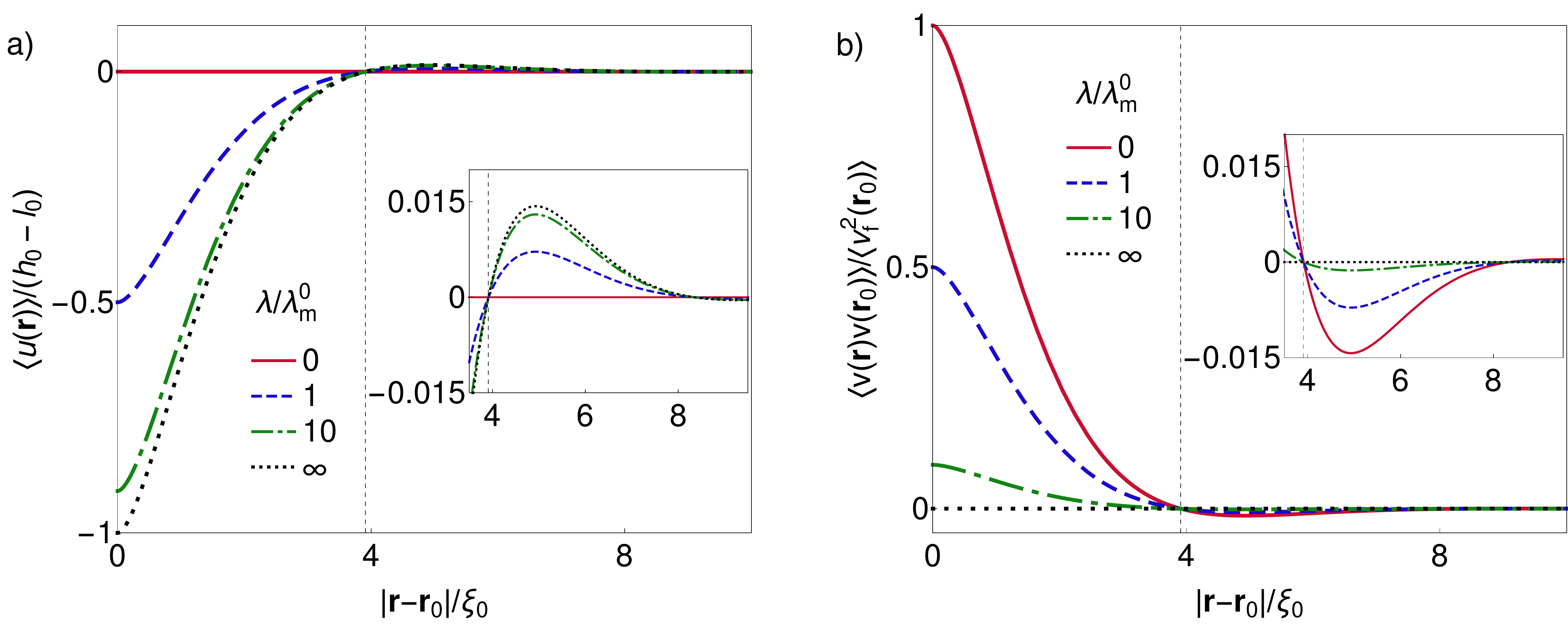}
 \caption{Spatial dependence of the static properties of the membrane for varying pinning stiffness $\lambda$. a) The mean shape given by eq. \ref{eq:Mean_profile} and b) the correlation function extracted from eq. \ref{eq:Correlation_from_the_pinning} show the same properties. The overshoots in the shape coincide with anticorrelations presented in the insets. Note that a more conventional parametrization of the mean shape in terms of height above the substrate is trivially obtained with $\langle h({\bf r}) \rangle=h_0+\langle u({\bf r}) \rangle$. Parameters: $\kappa=20 k_\text{B}T$, $\sigma = 10^{-20}  k_\text{B}T/\text{nm}^{2}$, $\gamma=3 \times \ 10^{-7}  k_\text{B}T/$nm$^{4}$, and $h_0-l_0=1$ nm.
  \label{fig:Mean_shape_and_correlation_with_the_pinning_site}
}
\end{figure}

Interestingly, following eqs. \ref{eq:system_of_equations_for_r_j_Greens_function} and \ref{eq:Mean_profile}, the correlation function  and the mean shape can also be expressed in terms of the correlation function for the free membrane
\begin{align}
 \label{eq:Mean_profile_rescaled_correlation_free}
 \langle u ({\bf r}) \rangle = -\mathcal K (h_0-l_0) \langle v_f ({\bf r}) v_f ({\bf r}_0) \rangle,
\end{align}
which emerges from the proportionality between the pinned- and the free-membrane correlation functions
\begin{align}
 \label{eq:Correlation_from_the_pinning}
 \frac{\langle v({\bf r}) v({\bf r}_0) \rangle}{\langle v_f ({\bf r}) v_f ({\bf r}_0) \rangle} & =  \frac{\langle v^{2}({\bf r}_0) \rangle}{\langle v^{2}_f ({\bf r}_0) \rangle} = \frac{\lambda_{\text{m}}}{\lambda + \lambda_{\text{m}}}.
\end{align}

This result clearly captures the interplay between the pinning stiffness $\lambda$ and the parameters of the membrane ($\sigma$ and $\lambda_{\text{m}}^{0}$) which are combined in $\lambda_{\text{m}}$.  If $\lambda\ll\lambda_{\text{m}}$, the pinning does not affect membrane fluctuations, whereas if $\lambda\gg\lambda_{\text{m}}$ fluctuations at the pinning are completely suppressed and small changes in $\lambda$ do not affect the system behavior. However, in the regime $\lambda\approx \lambda_{\text{m}}$ fluctuations can change noticeably, even for small change in the pinning stiffness (Fig. 3). Low tensed membranes will show such sensitivity if $\lambda\approx \lambda_{\text{m}}^{0}$ (large $\lambda/\sigma$ in Fig. 3b), while highly tensed membranes do so if $\lambda\gg \lambda_{\text{m}}^{0}$ (small $\lambda/\sigma$ in Fig. 3b).  Moreover, since the decay of correlations from the pinning site is independent of $h_0$ and $l_0$  (i. e., from the mean deformation), elastic properties of the pinning can be extracted directly from the change in the fluctuation amplitude between the pinned and the free states of the membrane.

\begin{figure}[t]
\centering
  \includegraphics[width=1\linewidth]{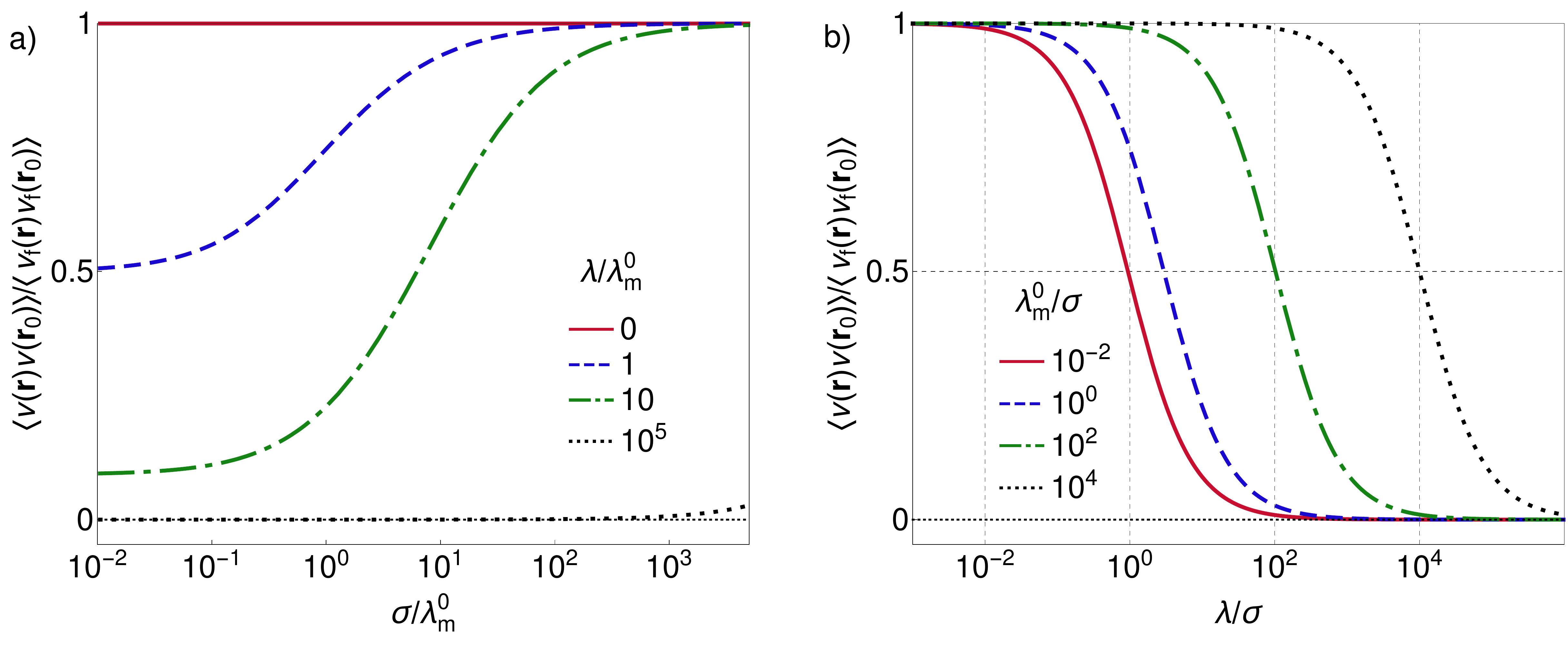}

\caption{ Effect of the pinning on the membrane fluctuations (eq. \ref{eq:Correlation_from_the_pinning}). a) Varying $\sigma$ and $\lambda$. b) Varying $\lambda$ and $\lambda_{\text{m}}^{0}$.
 \label{fig:pinnendmembrane}
}
\end{figure}

Another interesting relation is the one between the spatially-dependent mean square fluctuation amplitude and the square of the membrane shape 
\begin{align}
 \label{eq:FDRICM}
 \langle v({\bf r})^2 \rangle =   \frac{1}{\lambda_{\text{m}}}-\frac{ \langle u({\bf r}) \rangle^2}{\mathcal K (h_0-l_0)^2 }.
\end{align}
Both of these features can be measured using reflection interference contrast microscopy (RICM) with very high accuracy \cite{limozin2009}. Using very sparsely distributed pinnings, and allowing for independent measurements of $\lambda_{\text{m}}$ and $h_0 - l_0$,  stiffness of the pinning becomes the only unknown parameter, which can thus be extracted by comparing the shape and fluctuation profiles. So far the stiffness of the proteins was typically measured using atomic force microscopy, but outside of membrane environment, so this relation opens a possibility to extract mechanical properties of the pinning protein in its native environment. 

Actually, the existence of such a relation has been inferred in imaging of pinning sites using RICM \cite{smith2008, smith2010b}. In these studies,  the  suppression of membrane fluctuations was used to identify pinning sites that are of a lateral  dimension smaller than the optical resolution of the microscope,  which was possible because the correlation length of the membrane was similar or larger than the diffraction limit of the setup. Further development of this approach relies however on the understanding of the dependence of the correlation length of the pinned membrane on system parameters, as provided herein.  
%POTENTIALLY EXPAND WITH KHEYA

%********************************************************************************
%********************************************************************************
%********************************************************************************
%********************************************************************************

\newpage
\section{Effect of the membrane tension on the long-range behavior of the shape and correlation function}

 Both the mean shape and the correlations from the pinning site are proportional to the free membrane correlations. Hence, the decay length of the correlation function will be that of the free membrane correlation function, implying the insensitivity of the correlation length and the deformation range to the length and stiffness of the pinning.  Accordingly, dependent on various regimes (see SI section V for details), a power law and an oscillatory behavior are dominated by an exponential decay of a length  $\xi(\kappa, \sigma, \gamma)=\xi(\xi_0,\sigma/\lambda^{0}_m)$ identified through
\begin{equation}
   \langle v_f({r}) v_f(0) \rangle =g_f(r) \overset{r\rightarrow\infty}{\sim}  e^{- r/\xi(\xi_0,\sigma/\lambda^{0}_m)},
   \label{eq:Asymptotic_Green}
\end{equation}
where
\begin{equation}
\frac{ \xi}{\xi_0}=
    \begin{cases}
         \sqrt{2}  & \text{if } \sigma = 0,\\
\\
\left( \cos \left( \frac{1}{2} \arctan \left (\sqrt{\left(\frac{\lambda_{\text{m}}^{0}}{4\sigma}\right)^{2}-1}\right)\right) \right)^{-1}  & \text{if } 0 < \sigma < \frac{\lambda_{\text{m}}^{0}}{4},\\
\\
        1& \text{if } \sigma=\frac{\lambda_{\text{m}}^{0}}{4} ,\\
\\
        \left[\frac{4\sigma}{\lambda^{0}_m}\left(1 - \sqrt{1-\left(\frac{\lambda_{\text{m}}^{0}}{4\sigma}\right)^{2}}\right)\right]^{-1/2} & \text{if } \sigma> \frac{ \lambda_{\text{m}}^{0}}{4}.
    \end{cases}
    \label{eq:limitgf}
\end{equation}

Remarkably, increasing tension does not necessarily induce longer range height correlations. Instead, when bending dominates,  small amounts of tension ($\sigma < \frac{1}{4}\lambda_{\text{m}}^{0}$) actually reduce the decay length of correlations (Fig. 4).   In this regime, the membrane shape and correlation function exhibit an overshoot / anti-correlations of the long range limit immediately after the pinning (Fig. 2), followed by an oscillatory behavior  within an exponentially decaying envelope (Eq. SI-V.8). Similarly to systems that are governed only by bending and tension (no non-specific potential), the tension here flattens the membrane so that the spatial correlations decrease, due to changes in curvature which decay faster as the distance from the inclusion increases. Specifically, as the tension increases toward the critical value of $\sigma_c=\lambda_{\text{m}}^{0}/4$, the amplitude of the oscillations decreases. When the tension reaches $\sigma_c$, the oscillations are completely flattened, and the system enters a tension dominated regime. Now, coupling to the non-specific potential induces a slow, purely exponential decay of the shape and the correlations (SI-V.7). In this case, the larger the tension, the longer the range of the deformation and the correlation function, simply because of the increase in the energy penalty for large curvatures in a nonspecific potential. However, only when the tension reaches $\sigma=5\lambda_{\text{m}}^{0}/16$ the correlation length becomes longer than that of a tensionless free membrane.

\begin{figure} [h!]
\centering
\includegraphics[width=0.45\linewidth]{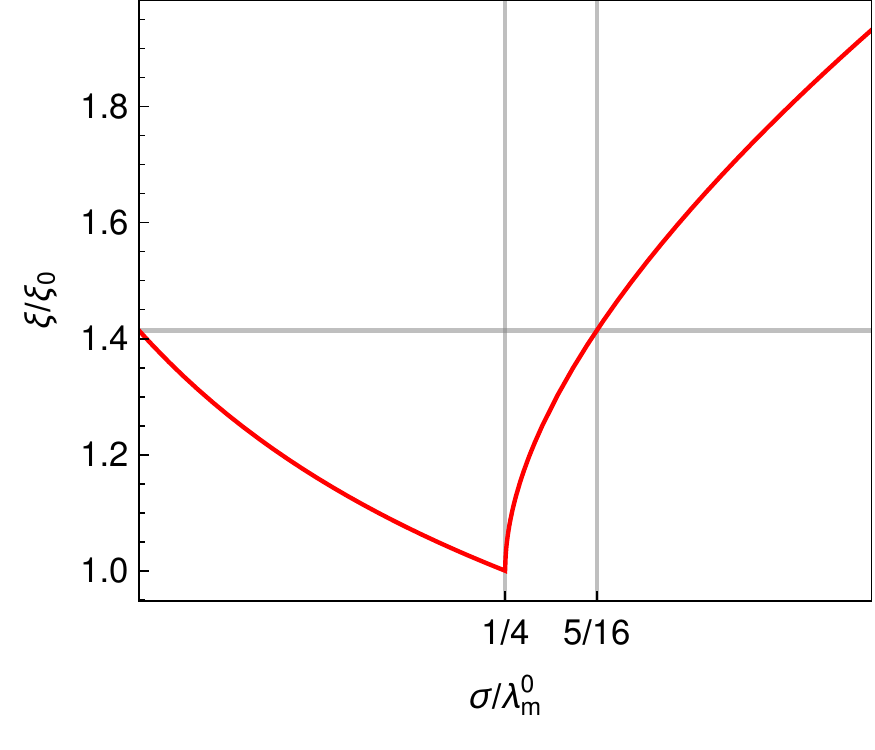}
\caption{Correlation decay length $\xi(\xi_0,\sigma/\lambda^{0}_m)$ in the asymptotic limit $r\to\infty$ (eq. (\ref{eq:limitgf})). The dark grey lines mark the value $\sigma = (5/16) \lambda_{\text{m}}^{0}$, beyond which an increase in tension results in longer range correlations than in the tensionless case.}
\label{fig:AsymptoticDecayStrength}
\end{figure}

Notably, the mean shape and correlations (and their derivatives with respect to the spatial coordinate $r$) are continuous functions of $\sigma$, even at $\sigma_c$, and no actual singularity appears in the system at the crossover between the bending and tension dominated regimes.

%********************************************************************************
%********************************************************************************
%********************************************************************************
%********************************************************************************
%********************************************************************************

\section{Membrane-mediated interactions between two pinnings}

Equations \ref{eq:Asymptotic_Green} and \ref{eq:limitgf} are significant in the context of interactions between pinnings on the membrane separated by a relative distance $x$. Following previous work \cite{schmidt2012} the interaction energy between two pinnings is 
\begin{align}
    \mathcal V_{2}(x) =  \frac{\mathcal K (l_0 - h_0)^2}{1+\mathcal K g_f(x)} + \frac{1}{2}\ln  \left (  1 - \left[\mathcal K  g_f(x)  \right] ^2 \right) ,
    \label{eq:potential}
\end{align}
where the first term is the deformation energy stored in the system with two bonds and the second term is the entropic cost associated with the suppression of fluctuations (see SI Section IV for details of the calculation). Terms which are independent on the relative distance between the two pinnings are omitted, since they drop out in the calculation of the force between two pinnings $\mathcal F_2(x)=-\partial\mathcal V_2(x)/\partial x$, which becomes

\begin{align}
    \mathcal F_{2}(x) =  \frac{\mathcal K^2 (l_0 - h_0)^2 g'_f(x)}{[1+\mathcal K g_f(x)]^2} + \frac{\mathcal K^{2} g_f(x) g'_f(x)}{1-\mathcal K^2 g_f(x)^2}.
    \label{eq:force}
\end{align} 
 Thus, the spatial dependence of the force is given by the correlation function of a free membrane at the relative distance $x$. The first term on the right hand side of eq. \ref{eq:force} can be associated with the force that emerges due to the membrane deformation, while the second term is the force arising from the suppression of membrane fluctuations in a spatially-dependent manner. If the pinning deforms the membrane ($h_0\neq l_0$), the deformation term determines the long-range behaviour of the force, as it decays two times slower than the fluctuation term (Fig. 5). Namely, the deformation term is proportional to $g'_f(x)$, which decays exponentially, and independent of the amount of the deformation in the system, while the fluctuation term, being proportional to $g_f(x) g'_f(x)$, decays exponentially but twice as fast (Fig. 5a). The deformation term typically dominates closer to the pinning as well (Fig. 5). However, if $h_0 \simeq l_0$, fluctuation forces dominate, in which case the decay length of the force is halved in comparison  to the case of a deformed membrane. This means that even if the protein does not affect the membrane shape ($h_0 = l_0$), significant force may emerge and potentially lead to the agglomeration of pinning sites, as suggested by simulations of a membrane containing many pinnings, described by the same Hamiltonian  \cite{reister2008, fenzbihr2011, speck2010, bihr2015}. While only limited understanding of the conditions necessary for the formation of domains is available at the moment, access to eq. \ref{eq:force} sets the foundation of the calculation of critical parameters which are necessary for the process of agglomeration.

\begin{figure}[h]
\centering
  \includegraphics[width=1\linewidth]{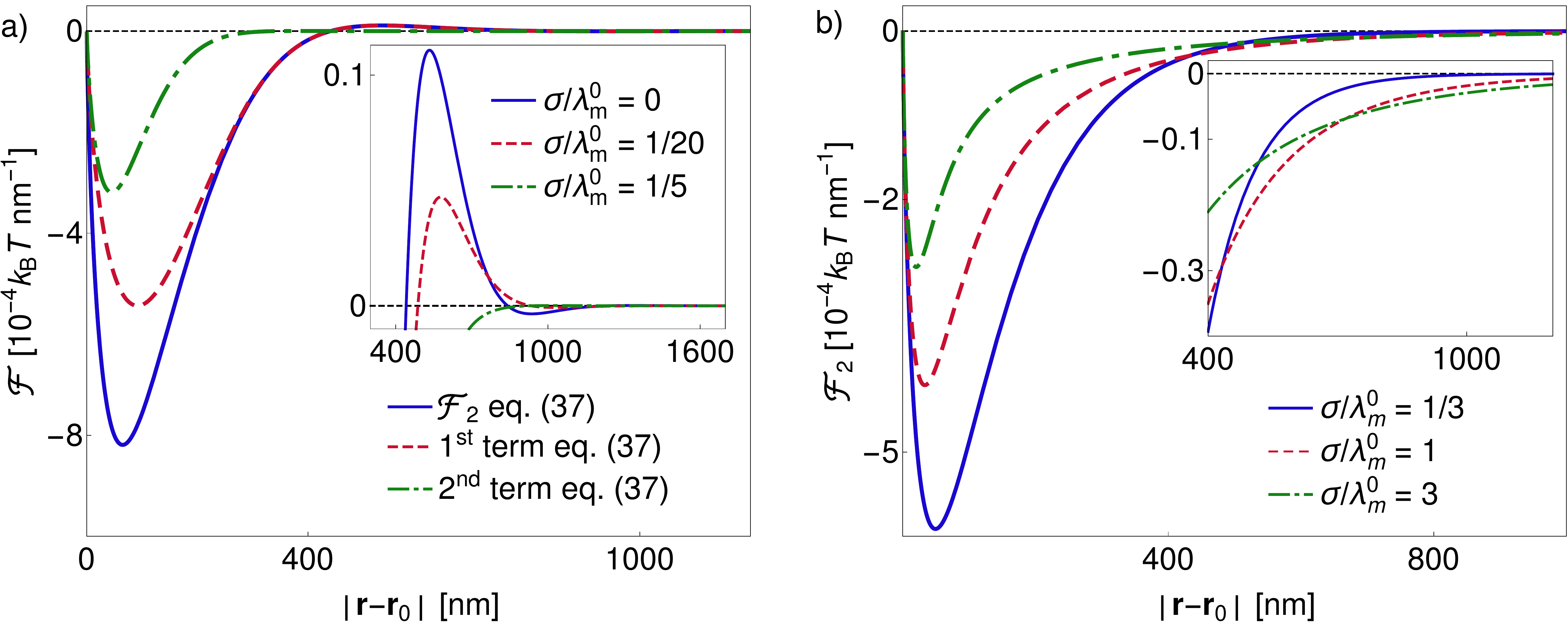}
\caption{Force between two pinnings. a) Bending dominated regime ($\sigma/\lambda_{\text{m}}^{0} < 1/4$) is shown. For this specific set of parameters, the deformation and the fluctuation contributions to the force are comparable. Both contributions oscillate around zero (inset), but the fluctuation part decays two times faster. b) The tension dominated regime ($\sigma/\lambda_{\text{m}}^{0} > 1/4$). As we increase the tension, the range of the force increases. Parameters: $\lambda = 0.75\times 10^{-2} k_\text{B}T/\text{nm}^{2}$, $\kappa = 20 k_\text{B}T$, $\gamma = 3.125\times 10^{-7} k_\text{B}T/\text{nm}^{4}$, $h_0-l_0=10$ nm.
 \label{fig:forces_between_pinnings}
}
\end{figure}

Based on the qualitative behaviour of the forces, we can recognize two regimes, namely the bending dominated ($\sigma < \sigma_c$) and the tension dominated regime ($\sigma > \sigma_c$). These regimes correspond to different regimes of the correlation function (see SI section V.). In the bending dominated regime, a  repulsive barrier appears in the force at distances of few membrane correlation lengths (Fig. 5a). Increasing tension, but staying under $\sigma_c$, flattens the barrier and the oscillating tail of the force (Fig. 5a - inset). This is contrasted by the tension dominated regime in which the repulsive barrier and the oscillating tail disappear and the long range forces are attractive (Fig. 5b). Moreover, the range of the force increases with tension (Fig. 5b - inset). In all cases the range of these weak interactions is of the order of 100 nm which is  nearly two orders of magnitude more than the direct protein-protein interactions. They are therefore considered long range, despite their universally-exponential nature.      

This exponential decay is contrasted by a body of work performed on forces between membrane inclusions in "bending only", or "tension only" systems for which the differential operator exhibits no scale. In the former case, the Green's function behaves as  $g_f(r) \sim  r^2 \log r^2$~ \cite{Goulian1993, Park1996, dommersnes1999}, and switching tension affects the power law nature of the decay \cite{weikl1998, evans2003, lin2011}. Because  the nonspecific potential introduces a length scale, the pinned membrane clearly delineates from these models for inclusions. However, it was recently proposed that a Hamiltonian, which is mathematically identical to that in eq. 1 can be used to model the inclusion of a protein with hydrophobic mismatch into a membrane \cite{bitbol2010}. Although the parameter range in which the linearized theory is valid could be more narrow than in the case of pinnings, the analogy of formalisms between the two problems, in principle, allows for the exploitation of the current results. Consequently, exponential decays should also appear in forces acting between membrane inclusions. However, these forces will have very different magnitudes and overall range.   

It is worth mentioning that so far we neglected the finite size of proteins. This is appropriate for sparse or immobile protein attachments (size of the attachment is still smaller than the correlation length of the membrane). When proteins approach within a few nanometers  separations between their surfaces, direct protein interactions will compete with the typically attractive membrane mediated interactions. The result of this competition at short range is non-universal,  and is most likely dominated by the direct contributions. Our hope is that the current approaches can be expanded to account for this case - either using the GF approach in analytic calculations, or using the expansions into relevant basis set for numerical simulations.

\section{Discussion and Conclusions}

In this paper, we studied the effect of a pinning on the statics of a membrane fluctuating in a harmonic nonspecific potential. We showed that the membrane and the pinning can be seen as two springs in series in the context of the energetics, as discussed previously \cite{fenz2017}. Hence, in the case when the length of the pinning does not coincide with the position of the undisturbed membrane in an effective potential, the deformation in the system depends on the effective spring constants of the pinning and the membrane (the later characterized by the inverse of the fluctuation amplitude in the absence of pinning). For stiff membranes, the pinning will extend its shape, while for stiff pinnings, membrane deformation will be considerable. However, since the lateral correlation length of the membrane is not affected by the pinning properties, the range of the deformation is independent of the pinning. This is very different to the effect of tension, which directly affects the correlation length, in a non-homogeneous fashion. 

The pinning, on the other hand, has a major effect on the membrane fluctuation amplitude,  which is an inverse function of the pinning stiffness. The correlation length  and the long range exponential behavior is, however, fully given by the correlation length of the free membrane. For small tensions, a pinning may induce short-range anticorrelations of fluctuations and an overshoot of the membrane shape. In this regime, the correlation length decreases with increasing tension. At high tensions, the correlation length increases, while the shape and the correlations continuously decay to their long-range limits. These correlations translate into long-range interactions between pinnings, which also decay exponentially. The forces associated with this interaction potential are stronger if the pinning displaced the membrane, however, even  in the absence of the deformation, the pinnings interact due to the suppression of fluctuations, analogously to Casimir forces. 

The results presented here open the possibility for differentiating between actively and passively pinned membranes in experiments, just by measuring the shape and fluctuations around a binding site, which can be either a single protein or a nanodomain, when the line tension remains small. Violation of the  relationship (eq. \ref{eq:Mena_shape_rescaled_correlation} - \ref{eq:FDRICM}) between the correlation functions and the shape provided in Section IV could be taken as a notion of activity. Moreover, in passive systems with small non-linear effects, exploiting the same relations could provide the foundation for the measurement of the stiffness of proteins in their natural membrane environment. The here-proposed  models should be suitable for analysis of data obtained using interferometric methods, or in conjunction  with atomic force microscopy of membrane-protein interactions, where vesicles are used as soft probes.

Given that membranes, locally pinned by proteins or macromolecular assemblies, are indeed ubiquitous in nature, a toolbox developed herein consisting of mode-coupling coefficients, orthonormal modes and the Green's function of the system is highly useful for future theoretical studies of membranes which aim to elucidate the interplay between the membrane elasticity and the forces transmitted by the proteins in the biological context.  We may anticipate that the Green's function approach may be the method of choice for analytic modeling, however,  normal modes and the mode coupling coefficients for the plane waves may be particularly useful in the context of numerical calculations. 
Of course, the equivalence of all three approaches can be stated by construction. Nevertheless, in terms of results presented herein,  GF and plane wave approaches give exactly the same representation of the mean shape (eq. \ref{eq:Mean_profile} vs eq. \ref{eq:Mean_profile_pmane_waves}) and the correlations (eq. \ref{eq:Spatial_correlations_Greens} vs eq. \ref{eq:Spatial_correlations_plane_waves}), while the normal modes give an alternative, but numerically identical representation (eq. \ref{eq:mean_shape_normal_modes} and eq. \ref{eq:Spatial_corelation_is_Greens_function_orthonormal_functions_expansion} for the mean shape and correlations, respectively).

 Besides studies in which membranes are used as probes for proteins binding during cell-cell and cell-substrate adhesion, or in the analysis of the interaction of the cytoskeleton with the plasma or nuclear membranes, which were in some cases based on the same Hamiltonian, other systems may benefit from the here developed tools and relations. In particular, as pointed out in the recent work of \cite{bitbol2010}, the same Hamiltonian could be used in studies of the interactions between membrane inclusions \cite{muller2005,sigurdsson2013,deserno2015}. However, since the energetics and the length scales of characteristic interactions are very different, non-linear  corrections may become important.  As there is a wealth of systems  where protein mediated pinning is important in the biological and biotechnological context, further developing a theory to account for the fluctuation dynamics of a permanently, but also stochastically pinned membranes appears as a natural and necessary extension of the current work, a task that we plan to undertake in our future work.

\vspace{5pt} $\bf{Acknowledgments}$: A.-S.S. D.S. H.S and J.A.J were funded by ERC Starting Grant MembranesAct 337283. J.A.J and A.-S.S were in part supported by Croatian Science Foundation research project CompSoLs MolFlex 8238. D.S. was member of the Research Training Group 1962 at the Friedrich-Alexander-Universität Erlangen-N\"urnberg.

\vspace{5pt} $\bf{Author\ Contributions}$: A.S.S and U.S conceived the study. A.S.S was in charge of overall direction and supervision. U.S.provided critical feedback and helped shape the research. D.S. obtained the mode-coupling coefficients. J.A.J. and D.S calculated the orthonormal modes. J.A.J. developed the Green's function approach. H.S. performed the asymptotic analysis. The force between two pinnings was calculated by H.S. and J.A.J.. A.S.S,  J.A.J., and H.S. wrote the manuscript with contributions from all authors. The authors declare that they have no competing interests. All data needed to evaluate the conclusions in the paper are present in the paper and/or the Supplementary Materials. All data and computer code for this study are available on request from the authors. 

\appendix
\section{Normal modes expansion}
\subsection{Solution of the eigenmode equation}

It remains to determine the normal modes $\psi_{j}$ given by eq. \ref{eq:Eigenequation_eigenfunctions}. By placing the pinning at the origin (${\bf r}_0=0$), the solution of eq. \ref{eq:Eigenequation_eigenfunctions} obeys radial symmetry with respect to the pinning site. Hence, the eigenmodes are a product of axial and radial functions, characterized by relevant mode numbers $m$ and $n$, respectively
\begin{align}
\psi_{nm} ({\bf r}) = R_{nm}(r) \mathrm e^{im\phi},
\end{align}
where $(r,\phi)$ are polar coordinates of the position ${\bf r}$. In this case, eq. \ref{eq:Eigenequation_eigenfunctions} takes the form
\begin{align}
\left [ \kappa \nabla^4 - \sigma \nabla^2 + \gamma +\lambda \delta({\bf r}) \right ] \psi_{nm} ({\bf r}) = E_{nm} \psi_{nm} ({\bf r}),
\label{eq:eigenvalue}
\end{align}
where $E_{nm}$ are the eigenvalues corresponding to modes $\lbrace{ n,m\rbrace}$. The square brackets on the left hand side enclose the energy operator which must be Hermitian (SI section II.A.1). 

The general solution of eq. \ref{eq:eigenvalue} emerges as a sum of Bessel functions (SI section II.B)
\begin{align}
R_{nm}(r) = & a_{nm} J_m(q_{nm} r) + b_{nm} Y_m(q_{nm} r) + c_{nm} K_m(Q_{nm} r)+ d_{nm} I_m(Q_{nm} r)\\\nonumber
\end{align}
with
\begin{align}
Q_{nm} = & \sqrt{q_{nm}^2 + \frac{\sigma}{\kappa}}.
\end{align}
Here, $J_m$ and $Y_m$ are Bessel functions of the first and second kind,  $K_m$ and $I_m$ are the modified Bessel functions of the first and second kind, respectively, and $a_{nm}$, $b_{nm}$, $c_{nm}$ and $d_{nm}$ are coefficients associated with the $n$ and $m$ mode numbers.

The corresponding eigenvalues in eq. \ref{eq:eigenvalue} are given by
\begin{align}
E_{nm} =\kappa q_{nm}^4+\sigma q_{nm}^2+\gamma,
\label{eq:eigenenergy}
\end{align}
and the general solution $R_{nm}(r)$  is specified by appropriate boundary conditions.

\textbf{Boundary condition 1} - 
$R_{nm}(r)$ stays finite when $r\to 0$: The Bessel functions of the first kind, $J_m$ and $I_m$, inherently fulfill this boundary condition ($J_0(0)=I_0(0)=1$ and $J_m(0)=I_m(0)=0$ for $m>0$). 
The remaining Bessel functions $Y_m$ and $K_m$ diverge for $r\to 0$. However, for $m=0$, both Bessel functions diverge logarithmically such that the sum $b_0Y_0(q_{nm}r)+c_0K_0(Q_{nm}r)$ stays finite with $c_0=2b_0/\pi$, while for $m>0$ such cancellation is not possible.  Consequently,
\begin{align}
\label{eq:Solution_after_finiteness_and_before_integral_vanishing} 
R_{nm}(r) =& a_{nm} J_m(q_{nm} r) + d_{nm} I_m(Q_{nm} r) + \delta_{m0} b_{nm} \left ( Y_m(q_{nm} r) + \frac{2}{\pi} K_m(Q_{nm} r) \right ),
\end{align}
where $\delta_{m0}$ is the Kronecker delta. The term multiplied by $\delta_{m0}$ is contributing only for $m=0$.

$ \quad \textbf{Boundary condition 2}$ - The integral of the eigenvalue equation \ref{eq:eigenvalue} over an infinitesimally small disk $D(\epsilon)$ centered at the pinning has to vanish,

\begin{align}
\int\limits_{D(\epsilon)} \mathrm d {\bf r} \left [ \kappa \nabla^4 - \sigma \nabla^2 + \gamma + \lambda \delta({\bf r}) \right ] \psi_{nm}({\bf r})= 0.
\end{align}

This boundary condition, often introduced around a delta function, is necessary to  ensure the finiteness of the membrane profile at the origin. With this imposed, the integration of the right hand side of the  eigenequation 5 vanishes in the relevant limit, and the limit is well defined. By extension, the integral of the left-hand side of the eigenequation 5 vanishes too (see SI section II.  for details).

By solving the integral for each mode, one obtains
\begin{align}
\label{eq:b0} 
b_{n0} = \Pi (q_{n0}) (a_{n0} + d_{n0}),
\end{align}
where
\begin{align}
\label{eq:Aq} 
\Pi (q_{n0}) = \frac{\lambda}{8\kappa (q_{n0}^2+\frac{\sigma}{2\kappa}) + \frac{\lambda}{\pi} \ln (1+ \frac{\sigma}{\kappa q_{n0}^{2}})}\ .
\end{align}

$ \quad  \textbf{Boundary conditions 3 and 4}$ - At the membrane edge, $r = P$, we have
\begin{align}
&R_{nm}(P)=0 
\label{eq:BC_at_the radius_a} \\
&\Delta R_{nm}(P)=0,
\label{eq:BC_at_the radius_b} 
\end{align}

where $\Delta$ denotes the Laplacian operator.
These boundary conditions arise in pair after imposing hermiticity of the operator in the eigenvalue equation  \ref{eq:eigenvalue} as shown in SI section II.

From eqs. \ref{eq:b0}-\ref{eq:BC_at_the radius_b} we obtain the asymptotic form of $R_{nm}(r)$ for a large membrane radius $P$ (SI section II.C.2)
\begin{align}
R_{nm}(r)  & \sim   a_{nm} \bigg \lbrace J_m(q_{nm} r) \left. + \delta_{m0} \left[ \Pi (q_{n0}) \left ( Y_{m}(q_{nm} r) + \frac{2}{\pi} K_{m}(Q_{nm} r) \right)\right] \right\rbrace,
\label{eq:Approx._of_the_eigenfunction_for_large_R} 
\end{align}
with
\begin{align}
q_{nm}\sim n\frac{\pi}{P}.
\label{eq:q_nm_asymptotic} 
\end{align}
This asymptotic form of  $q_{nm}$ emerges when $n\to\infty$ and membrane radius $P\to\infty$ as shown in SI-Section II.B-C.

$  \textbf{Normalization}$  of the solution of the eigenvalue problem (SI section II.C.2),  requires setting 
\begin{align}
a_{nm}= \frac{i^{m}}{\sqrt{ \left(1+\delta_{m0}\left(\Pi(n\Delta q)\right)^{2}\right)}}.
\label{eq:a_nm} 
\end{align}
Finally, by letting $P\to\infty$, $q_{nm}\to q \in \mathbb{R}$, the basis functions become $\psi_{m}({\bf r},q)$, and are given by (SI section III.C.3)
\begin{align}
&\psi_{m}({\bf r},q) =\frac{i^me^{im\phi}}{\sqrt{  \left(1+\delta_{m0}\left(\Pi(q)\right)^{2}\right)}}\left[ J_m(q r) + \delta_{m0} \Pi(q) \left ( Y_m(q r) + \frac{2}{\pi} K_m\left(\sqrt{q^{2}+\frac{\sigma}{\kappa}} r\right) \right) \right].
\label{eq:Eigenfunctions} 
\end{align}
Naturally, the orthogonality condition
\begin{align}
\int\limits_{\mathbb{R}^2}\mathrm{d}{\bf r} \psi_{m}({\bf r},q) \psi^*_{m'}({\bf r},q')= \frac{\delta(q-q')}{q} 2\pi \delta_{m,m'}
\label{eq:Orthonormality_of_the_basis_functions} 
\end{align}
is satisfied, and the profile of an infinite pinned membrane can be expanded in the basis functions $\psi_{m}({\bf r},q)$ as (SI section II.C.4)
\begin{align}
u({\bf r})=\frac{1}{2\pi} \sum^{\infty}_{m=-\infty} \int^{\infty}_{0}dq qU_m(q) \psi_{m}({\bf r},q),
\label{eq:Pinned_membrane_expansion_int_the_orthonormal_modes} 
\end{align} 
with
\begin{align}
U_m(q) = & 2\pi \int^{\infty}_0 \mathrm{d}r r  \ u_m(r) R^*_m(r, q),
\label{eq:Integral_transform_inverse_real} 
\end{align}
where
\begin{align}
u_m(r)= & \frac{1}{2\pi} \int_0^{2\pi} \mathrm{d} \phi \ u({\bf r}) e^{-im\phi}. 
\label{eq:radial_coeff_in_Fourier_exp_in_angle_coordiante} 
\end{align} 

For vanishing $\lambda$ (SI section II.C.5) the eigenmodes are given by the Bessel functions $J_m(qr)$ for all $m$, which is equivalent to a basis set constructed from plane waves in radial geometry, as demonstrated for a free membrane. 
For a non-vanishing $\lambda$, on the other hand, the pinning properties affect explicitly only the eigenmode with $m=0$. 

\subsection{Representing shape and fluctuations}
Expansion of the mean shape of the membrane pinned at ${\bf r}_0=0$ is given only by $m=0$ modes (SI section II.D.2):
\begin{align}
\langle u({\bf r}) \rangle = \lambda (l_0-h_0)\frac{1}{2\pi}\int^{\infty}_{0}dq q\frac{R_{0}(r,q)R^{*}_{0}(0,q)}{E_q}.
\label{eq:mean_shape_normal_modes}
\end{align}
At the pinning site ${\bf r}=0$
\begin{align}
\langle u(0) \rangle =& \lambda (l_0-h_0)\frac{1}{2\pi}\int^{\infty}_{0}dq q\frac{\vert \psi_{0}(0,q)\vert ^2}{E_q} = \frac{\lambda}{\lambda + \lambda_{\text{m}}}(l_0-h_0).
\label{eq:mean_shape_at_pinning_normal_modes} 
\end{align}
The correlation function is given by
\begin{align}
\langle v  ({\bf r}_1)  v ^* ({\bf r}_2) \rangle &= g({\bf r}_1\vert{\bf r}_2)= \frac{1}{2\pi}\sum_{m=-\infty}^{\infty} \int_0^{\infty} dq q \frac{\psi_m ({\bf r}_1,q)\psi^*_m ({\bf r}_2,q)}{E_q},
 \label{eq:Spatial_corelation_is_Greens_function_orthonormal_functions_expansion}
\end{align}
and the fluctuation amplitude by
\begin{align}
\langle v^2 ({\bf r}) \rangle &= \frac{1}{2\pi} \sum^{\infty}_{m=-\infty} \int^{\infty}_{0}dq q\frac{\vert \psi_{m}({\bf r},q)\vert ^2}{E_q}.
\label{eq:fluct_pinned_eigenfunction}
\end{align}
At the position of the pinning site
\begin{align}
\label{eq:fluct_pinned} 
& \langle v^2({0}) \rangle = \frac{1}{2\pi} \int\limits_0^\infty \mathrm d q \ \frac{q}{\kappa q^4 + \sigma q^2 + \gamma} \frac{ \left ( 8\kappa q^2 +4 \sigma \right )^2}{ \lambda^2 + \left [ 8\kappa q^2 +4\sigma +\frac{\lambda}{\pi} \ln \left ( 1+\sigma/(\kappa q^2) \right ) \right ]^2} = \frac{1}{\lambda +\lambda_{\text{m}}}, \nonumber\\
\end{align}
The last equality, which coincides with eq. \ref{eq:fluctuation_amplitude_pinned}, was checked numerically to the machine precision for an arbitrary tension, and analytically for $\sigma=0$.

\section{Plane wave expansion}

\subsection{Mode-coupling}

Relating the shape and the fluctuation amplitude to the properties of the free membrane should be also possible in the most commonly used plane wave expansion
\begin{align}
 u({\bf r}) = \frac{1}{(2\pi)^{2}}\int_{\mathbb{R}^2} d{\bf k} \ u({\bf k}) \mathrm e^{i{\bf k r}},
 \label{eq:planewaves_infinite}
\end{align}
where for the mean shape we find
\begin{align}
 \label{eq:membraneprofile_plane_waves}
 \langle u({\bf r}) \rangle = \frac{1}{(2\pi)^{2}}\int_{\mathbb{R}^2} d{\bf k} \langle u({\bf k}) \rangle \mathrm e^{i{\bf k r}},
\end{align}
and for the correlation function
\begin{align}
 \langle v({\bf r}) v({\bf r}') \rangle &=  \frac{1}{(2\pi)^{4}}\int_{\mathbb{R}^2} d{\bf k}\int_{\mathbb{R}^2} d{\bf k}' \langle u({\bf k})u({{\bf k}^\prime}) \rangle \mathrm e^{i{\bf k r}} \mathrm e^{i{\bf k}^\prime {\bf r}'} - \frac{1}{(2\pi)^{4}}\int_{\mathbb{R}^2} d{\bf k} \int_{\mathbb{R}^2} d{\bf k}' \langle u({\bf k}) \rangle\langle u({{\bf k}^\prime}) \rangle \mathrm e^{i{\bf k r}} \mathrm e^{i{\bf k}^\prime {\bf r}'}.
\label{eq:latcorrfct_pinned}
\end{align} 
 The disadvantage of this approach is the coupling of the modes, giving rise to expansion coefficients $\langle u({\bf k})u({{\bf k}^\prime}) \rangle$ that have so far not been calculated explicitly.
 
As previously discussed \cite{schmidt2012}, the amplitudes $\langle u({\bf k}) \rangle$ and the mode coupling coefficients $\langle u({\bf k}) u({\bf k}') \rangle $ are defined as
\begin{align}
 \langle u({\bf k}) \rangle & \equiv \frac{1}{\mathcal Z} \int \mathcal D [ u] \ u({\bf k}) \exp \left [ -\mathcal H \right ], \nonumber \\
 \langle u({\bf k}) u({\bf k}') \rangle & \equiv \frac{1}{\mathcal Z} \int \mathcal D [ u] \ u({\bf k}) u({\bf k}') \exp \left [ -\mathcal H \right ],
 \label{eq:modecoupling_def}
\end{align}
with $ \mathcal Z $ being the partition function
\begin{align}
 \mathcal Z & = \int \mathcal D [ u] \  \exp \left [ -\mathcal H \right ].
\end{align}

Treating identities in eq. \ref{eq:modecoupling_def} as Gaussian integrals (SI section III), gives
\begin{align}
 \label{eq:mean_plane_waves_coeff}
 \langle u({\bf k}) \rangle = & -\frac{\lambda \lambda_{\text{m}}}{\lambda + \lambda_{\text{m}}} (h_0-l_0) \ \frac{e^{-i{\bf k}{\bf r}_0}}{\kappa k^4 + \sigma k^2 + \gamma}, \\
  \label{eq:modecoupling}
 \langle u({\bf k})u({\bf k}') \rangle = & \frac{\delta ({\bf k}+{\bf k}^\prime) }{\kappa k^4 + \sigma k^2 + \gamma} + \langle u({\bf k}) \rangle \langle u({\bf k}') \rangle -\frac{\lambda \lambda_{\text{m}}}{\lambda + \lambda_{\text{m}}} \frac{e^{-i{\bf k}{\bf r}_0}}{\kappa k^4 + \sigma k^2 + \gamma} \frac{e^{-i{\bf k}'{\bf r}_0}}{\kappa k^{\prime 4} + \sigma k^{\prime 2} + \gamma}.
\end{align}

\subsection{Representing shape and fluctuations}
Combining eqs. (\ref{eq:membraneprofile_plane_waves}) and (\ref{eq:mean_plane_waves_coeff}) we obtain the mean shape for a pinned membrane 
\begin{align}
 \label{eq:Mean_profile_pmane_waves}
 \langle u({\bf r}) \rangle = &\frac{\lambda \lambda_{\text{m}}}{\lambda + \lambda_{\text{m}}} (l_0-h_0) g_f({\bf r}-{\bf r}_0).
\end{align}

By combining eqs. (\ref{eq:latcorrfct_pinned}), (\ref{eq:mean_plane_waves_coeff}) and (\ref{eq:modecoupling}), we obtain for the spatial correlations
\begin{align}
 \langle &v({\bf r}) v({\bf r}') \rangle = g_f({\bf r}-{\bf r}')-\frac{\lambda\lambda_{\text{m}}}{\lambda+\lambda_{\text{m}}}g_f({\bf r}-{\bf r}_0)g_f({\bf r}_0-{\bf r}')
\label{eq:Spatial_correlations_plane_waves}
\end{align}
and for the fluctuation amplitude (${\bf r}={\bf r}'$)
\begin{align}
 \langle &v^{2}({\bf r}) \rangle = \frac{1}{\lambda_{\text{m}}}-\frac{\lambda\lambda_{\text{m}}}{\lambda+\lambda_{\text{m}}}g_f^{2}({\bf r}-{\bf r}_0).
\label{eq:Fluctuation_amplitude_plane_waves}
\end{align}
We have therefore independently derived the same result as with the Green's function approach (eqs. \ref{eq:Mean_profile} and \ref{eq:Spatial_correlations_Greens}).

\section*{Supporting Citations}

References \cite{abramowitz,baddour2009,NIST2010} appear in the Supporting Material.

\bibliography{Bibliography}

\end{document}